\documentclass[11pt,letterpaper]{article}
\usepackage[margin=1in]{geometry}
\usepackage[table]{xcolor}
\usepackage[T1]{fontenc}
\usepackage[utf8]{inputenc}
\usepackage{lmodern}
\usepackage[numbers,square,sort&compress]{natbib}
\usepackage{graphicx,booktabs,microtype,placeins,needspace,tabularx,float}
\usepackage[font=small,labelfont=bf]{caption}
\usepackage[hidelinks]{hyperref}
\hypersetup{pdftitle={Agentic RAG Evaluation: Budget Allocation Across Questions, Trajectories, and Reads},
  pdfauthor={Jingjie Ning, Xueqi Li, Yibo Kong},
  pdfsubject={Evaluation budget allocation in agentic RAG},
  pdfkeywords={Agentic RAG, LLM evaluation, evaluation budget allocation, variance decomposition, nested sampling, statistical precision}}
\usepackage{xurl}
\newcommand{\doi}[1]{\href{https://doi.org/#1}{\mbox{doi \nolinkurl{#1}}}}
\title{\bfseries Agentic RAG Evaluation: Budget Allocation Across Questions, Trajectories, and Reads}
\definecolor{AuthorLink}{HTML}{4F73B8}
\author{%
\textbf{Jingjie Ning}\textsuperscript{1}\thanks{Corresponding author.}\quad
\textbf{Xueqi Li}\textsuperscript{1}\quad
\textbf{Yibo Kong}\textsuperscript{1}\\[4pt]
{\normalfont\small \textsuperscript{1}Carnegie Mellon University}\\[2pt]
{\normalfont\footnotesize
\{\href{mailto:jening@cs.cmu.edu}{\textcolor{AuthorLink}{jening}},\,
\href{mailto:xueqil@cs.cmu.edu}{\textcolor{AuthorLink}{xueqil}},\,
\href{mailto:yibok@cs.cmu.edu}{\textcolor{AuthorLink}{yibok}}\}@cs.cmu.edu}}
\date{}
\definecolor{BrandNavy}{HTML}{4F587D}
\definecolor{BrandPink}{HTML}{C68DC0}
\definecolor{BrandCyan}{HTML}{C2E0EE}
\definecolor{BrandPurple}{HTML}{776B97}
\definecolor{BrandLavender}{HTML}{DBC8ED}
\begin{document}
\maketitle
\begin{abstract}
Evaluation budgets in agentic retrieval-augmented generation span questions, search trajectories, and repeated answers. We measure allocation precision, reading efficiency, and cost boundaries using a retrieval-feedback comparison on HotpotQA and MuSiQue. At 34.14--34.39M model tokens, broader question coverage lowers standard error by 33\% versus five reads and 12.6\% versus three trajectories. Archived nested and Q-only forecasts predict these allocations within 4.0\% and 3.5\%, respectively. Depth subsets establish no clear forecasting advantage beyond the two-trajectory audit. One-read variance penalties relative to the fitted optimum at the same token budget are 0--9.9\%, with substantial Pro uncertainty. Under recorded model fees, more questions beat more trajectories at search prices of \$0--1 per 1,000 requests; question-versus-read fee rankings remain unresolved. Temperature zero cuts answer disagreement from 14.3\% to 3.4\% while comparison precision stays similar.
\end{abstract}

\begin{figure}[H]
\centering
\includegraphics[width=\linewidth]{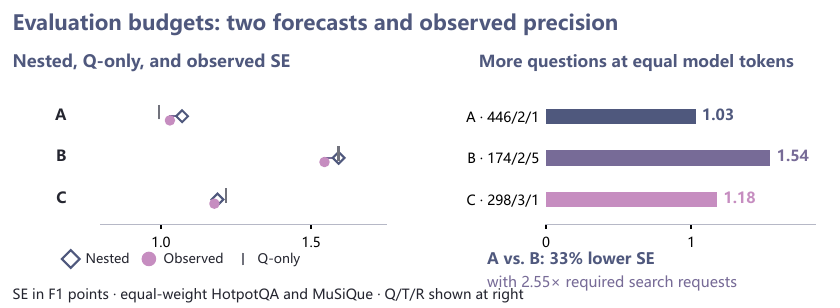}
\caption{Budget allocation and two audit-based forecasts. Nested and Q-only predictions both approximate measured standard errors (SEs, F1 points) in A/B/C; their B predictions coincide by construction. A improves precision at matched model-token cost while requiring more searches. Figure~\ref{fig:allocation} tests trajectory and reader scaling.}
\label{fig:teaser}
\end{figure}

\section{Introduction}

Agentic retrieval-augmented generation (RAG) evaluations distribute a budget across questions, search trajectories, and final answers \citep{rag}. Variation at each level affects the precision of a policy comparison. An evaluation can spend more on question coverage or repeated generations, with different consequences for both uncertainty and cost.

Broad topic coverage and variance-based sample design have established foundations in information retrieval \citep{sanderson2005evaluation,webber2008power,sakai2016topics} and generalizability theory \citep{bodoff2007test,urbano2013reliability}. Agentic RAG adds distinct planning, retrieval, and reading costs. Applying these principles requires measuring complete costs and checking how well an audit predicts precision as trajectory and read counts change.

We study this tradeoff through retrieval-feedback comparisons on HotpotQA and MuSiQue, with four queries and five reader documents per policy. A prior variance audit informs three allocations at nearly equal model-token cost. We measure realized precision and complete call costs, and test forecasts against saved trajectory and reader subsets. Model and temperature controls assess reading efficiency.

Broader question coverage yields 33\% lower standard error than five reads at matched model-token cost (Fig.~\ref{fig:teaser}), while fee comparisons depend on caching and search charges. Single-read evaluation is efficient in the tested Flash workflow; the Pro allocation remains uncertain. Q-only scaling predicts the initial allocation precision accurately, and deeper subsets establish no clear extrapolation advantage for the nested model. The study provides measured budget comparisons and empirical guidance on their sampling and cost boundaries, with scored outcomes and standalone analyses in the accompanying artifact.

\section{Related work}

\textbf{Evaluation design and reliability.} Sanderson and Zobel \citep{sanderson2005evaluation} and Webber et al.~\citep{webber2008power} show how broader topic coverage can improve comparisons at fixed relevance-assessment effort. Sakai \citep{sakai2016topics} designs topic-set sizes from power and interval-width targets. Bodoff and Li \citep{bodoff2007test} apply generalizability theory and derive cost-based allocation rules; Urbano et al.~\citep{urbano2013reliability} document the sensitivity of D-study predictions to the fitting sample. These works supply the statistical framework and motivate reporting prediction error, uncertainty, and a simple scaling baseline.

\textbf{Repeated model evaluation.} Miller \citep{miller2024errorbars} separates question and answer variation, develops paired inference, and explains why lowering temperature changes the evaluated behavior. IR testing and reporting studies \citep{urbano2019irtests,sakai2016power,fuhr2017eval}, repeated-run analyses \citep{reimers2017distributions,bouthillier2021variance}, and paired NLP evaluation \citep{dror,showwork} support our question-level analysis. Same Agent, Different Answers \citep{ning2026answerchurn} measures answer compatibility under a fixed single-turn policy as corpus access expands. Our outcome is the precision of a paired four-query policy comparison under nested repetitions.

\textbf{Agentic retrieval and resource allocation.} RAG \citep{rag} and DeepResearchGym \citep{coelho2025deepresearchgym} supply the task and access setting. Corpus--generator trade-offs characterize retrieval access as an inference resource \citep{ning2026lessllm}. Search logs document query repetition and lexical links to retrieved evidence \citep{ning2026searchwild}. Relevance feedback \citep{relevancemodel,prfpitfalls}, query expansion \citep{hyde,query2doc}, and interleaved search \citep{react,ircot} motivate the policy comparison. Evidence-utility diagnostics \citep{bridgeevidence,searchdiagnosis} distinguish retrieval quality from downstream benefit. RRF \citep{cormack2009rrf}, setwise retrieval \citep{rubric4setwise}, and context order \citep{lostmiddle} motivate explicit final-packet selection. Our focus is the cost and precision of repeated policy evaluation.

\section{Study design}

\subsection{Paired search policies}

Open plans four queries from the question and query history before retrieval. Adaptive plans queries two through four using earlier results. Each policy pair shares its first query and search result. The final reader answers from five documents. Our case outcome is the answer-F1 contrast $\Delta_{\rm FB}=\mathrm{F1}_{\rm Adaptive}-\mathrm{F1}_{\rm Open}$; the evaluation target is its standard error.

\begin{figure}[H]
\centering
\includegraphics[width=\linewidth]{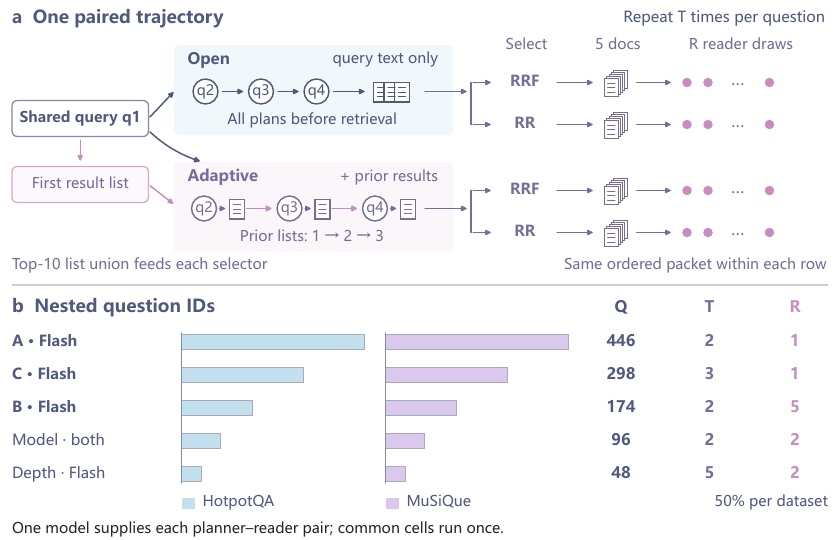}
\caption{Paired policies and nested question panels. RRF denotes reciprocal rank fusion and RR round robin. Repeated reads keep each packet fixed; identical packets share calls within a policy, trajectory, and draw.}
\label{fig:design}
\end{figure}

We use FineWeb through DeepResearchGym \citep{fineweb,coelho2025deepresearchgym}: seven partitions, ten results per query, and at most five 5,000-character document prefixes per observation or final packet. RRF sums $1/(60+r)$ across distinct query lists; RR cycles by rank across lists, skipping duplicates. Both select five records from a common pool. Appendix~\ref{app:cross} gives exact selector definitions.

Planner and reader use non-thinking requests, temperature 0.7, and a 256-token output cap. The model routes are DeepSeek-V4.1-Flash (\texttt{deepseek-flash}) and DeepSeek-V4-Pro-0813 (\texttt{deepseek-v4-pro}), verified against returned identifiers and provider documentation.\footnote{\url{https://api-docs.deepseek.com/quick_start/pricing/}, accessed October 1, 2026.} An immutable model or index snapshot is unavailable. Exact prompts and recorded serving settings are in Appendix~\ref{app:protocol}.

\subsection{Evaluation panels and sampling allocations}

The prior audit contains 96 questions, two trajectories per question, and five reads per fixed evidence packet. A separate 446-question panel supplies the three allocation designs in Fig.~\ref{fig:design}; its IDs and MuSiQue components exclude the audit. The new executions preserve equal HotpotQA/MuSiQue weights. A seeded shuffle selects nested subsets: A uses $446/2/1$, B $174/2/5$, and C $298/3/1$ questions/trajectories/reads.

Configurations, IDs, the primary selector-average contrast, and the 34.35M model-token target were fixed before these calls. Shared cells execute once across allocation analyses; all planned failures retain zero scores. The separate model control uses 96 questions with $T=2,R=2$, and the depth control uses 48 with $T=5,R=2$. The extension contains 12,824 scheduled cells. A same-packet temperature control on the prior 96 questions provides three reader draws at each temperature.

\FloatBarrier
\section{Estimation and budget accounting}

\subsection{Variance estimation and precision forecasting}

We use answer-normalized token F1 and exact match for HotpotQA \citep{hotpotqa} and MuSiQue \citep{musique}. Each question averages its trajectories and reads; the primary contrast also averages RRF and RR. Standard errors combine within-dataset variances with equal dataset weights. Primary paired 95\% percentile confidence intervals resample complete questions within datasets 10,000 times \citep{dror,miller2024errorbars}.

For dataset $d$, question $q$, trajectory $t$, read $r$, and selector $s$, let $D_{dqtr}^{(s)}$ denote the policy difference. Current component estimates fit separate dataset means and use within-dataset question mean squares with $Q-2$ degrees of freedom. Balanced nested moments yield
\begin{equation}
V(Q,T,R)\approx\frac{\sigma_Q^2}{Q}+\frac{\sigma_T^2}{QT}+\frac{\sigma_R^2}{QTR}.
\label{eq:variance}
\end{equation}
Second-moment estimation accommodates bounded, non-Gaussian F1 scores. Question bootstraps preserve their bounded distributions and nested dependence. Negative component estimates are truncated at zero. Figure~\ref{fig:variance} shows \emph{single-cell component shares}; Eq.~\ref{eq:variance} applies the distinct weights needed for the \emph{variance of the mean}. Small trajectory counts can still make the component estimates unstable.

For prediction, both models use the archived 96-question audit. Its parameters and Eq.~\ref{eq:variance} were fixed before the extension; numerical forecasts were computed retrospectively, with no parameters fitted or selected using the new outcomes. The nested model varies $Q,T,R$; the Q-only baseline retains the audit's per-question variance and scales SE by $1/\sqrt Q$. A/B/C share cells and form one allocation check.

We additionally enumerate every trajectory subset and every per-trajectory read subset in the 48-question depth panel ($T=1,\ldots,5$, $R=1,2$), and every read subset in B ($T=2$, $R=1,\ldots,5$). Empirical SE is the square root of the average estimator variance across subsets. All subsets receive equal weight; answers remain separate across subsets. We compare predictions using mean absolute error (MAE) across configurations and separately across $T>2$. Shared question bootstraps preserve overlap between configurations. Their 95\% intervals condition on recorded trajectories and omit prior-audit parameter uncertainty.

\subsection{Reader calls and one-read efficiency loss}

One draw under one selector costs up to two reader calls for the Open/Adaptive pair; averaging two selectors costs up to four. Actual totals incorporate packet reuse and failures. Let $c_T$ be paired planning tokens and $c_R$ tokens per paired reader draw for the chosen selector set. Holding $T$ fixed gives the D-study objective \citep{bodoff2007test}
\[
J_T(R)=(T\sigma_Q^2+\sigma_T^2+\sigma_R^2/R)(c_T+Rc_R).
\]
We report $100[J_T(1)/\min_R J_T(R)-1]$ for $R=1,\ldots,20$: excess estimator variance at the same model-token budget. The fixed-$Q$ calculation includes the unchanged question term and retains at least one trajectory (Appendix~\ref{app:efficiency}). Its free allocation level is continuous. Stratified question bootstraps jointly refit variance and cost 3,000 times. All intervals and descriptive loss ranges use the 2.5th--97.5th bootstrap percentiles (central 95\%).

We additionally compare the recorded one-read and two-read estimates directly on identical questions and trajectories, using empirical $\mathrm{SE}^2\times\mathrm{tokens}$. Each of the two single-read batches is checked separately, then their products are averaged. This comparison uses no nested component model. Its bootstrap retains paired questions, both read batches, and their observed costs.

\subsection{Cost accounting and reader temperature}

Model totals include planning and reading inputs and outputs. Paired accounting counts the shared first planner call once. For each allocation, we count both logical search operations and unique requests under the configured exact-query cache. Search-price sensitivity uses $C_j(p)=C_j^{\rm model}+pN_j$, where $p$ is USD per request. Model fees use recorded cache discounts and tariffs; price curves hold these fees fixed. An independent equal-dollar experiment is outside the reported evidence.

For temperature, the one-read SE summary is the square root of the average of three separately computed single-batch estimator variances. Each batch uses one read per packet. This keeps single-read precision distinct from the three-read mean. Temperature changes the operating configuration being evaluated \citep{miller2024errorbars}.

\section{Results}

\subsection{Precision forecasts across sampling configurations}

A/B/C have measured SEs of 1.03/1.54/1.18 F1 points. The archived nested model predicts 1.07/1.59/1.19, within 4.0\%; Q-only predicts 0.99/1.59/1.22, within 3.5\%. B retains the audit's $T=2,R=5$, so both predictions are identical by construction. Adjusting dataset means in the old audit changes predictions by less than 0.7\%.

\begin{figure}[H]
\centering
\includegraphics[width=\linewidth]{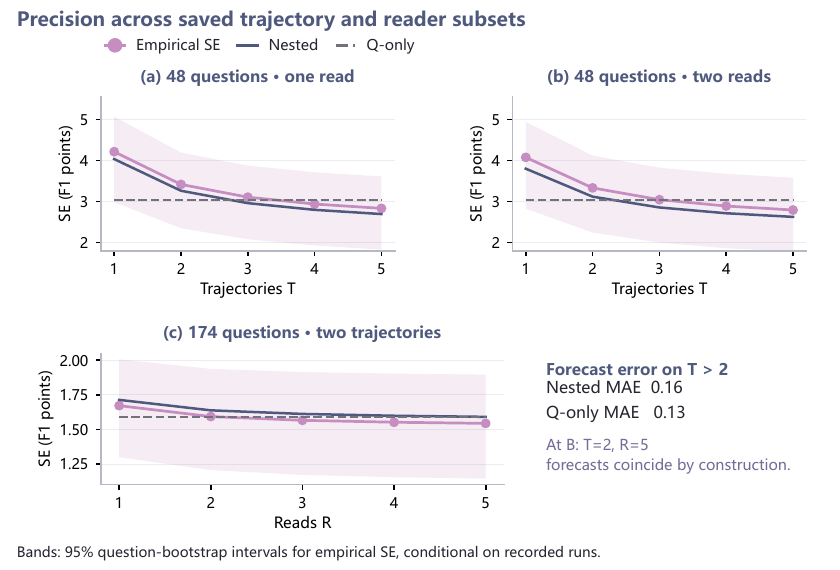}
\caption{Forecasts across saved subsets. (a,b) All trajectory and per-trajectory reader subsets on 48 questions. (c) All reader subsets in B. Empirical SE averages estimator variances before taking the square root. Bands show 95\% question-bootstrap intervals (2.5th/97.5th percentiles), conditional on recorded runs. Forecasts use fixed audit parameters.}
\label{fig:allocation}
\end{figure}

Figure~\ref{fig:allocation} probes changes in $T$ and $R$ directly. On 48 questions with one read, empirical SE falls from 4.21 at $T=1$ to 2.83 at $T=5$. Across both read settings, the nested model's MAE is 0.18 versus Q-only's 0.37 points; the paired MAE advantage is 0.19 [$-0.06$,0.36]. This gain comes from $T=1,2$. For $T=3,4,5$, errors are 0.16 versus 0.13, with an advantage interval of [$-0.25$,0.25]. Thus the two-trajectory audit establishes no reliable forecasting advantage when extrapolating $T$. In B, empirical SE decreases from 1.67 to 1.54 as $R$ grows from one to five; nested and Q-only MAEs are 0.045 and 0.039.

\subsection{Question coverage and evaluation cost}

The broad A allocation gives a feedback contrast of +1.66 [$-0.27$,+3.65] F1 points. A/B/C consume 34.20, 34.14, and 34.39M model tokens, a difference below 0.8\%. A reduces measured SE by 33.3\% relative to B and 12.6\% relative to C; both SE-ratio intervals favor A. This is a measured agentic-RAG instance of the broad-versus-deep allocation principle \citep{sanderson2005evaluation,webber2008power}. The comparison includes the reader and output tokens; Adaptive/Open's complete model-token ratio in A is 3.68.

\begin{figure}[H]
\centering
\includegraphics[width=\linewidth]{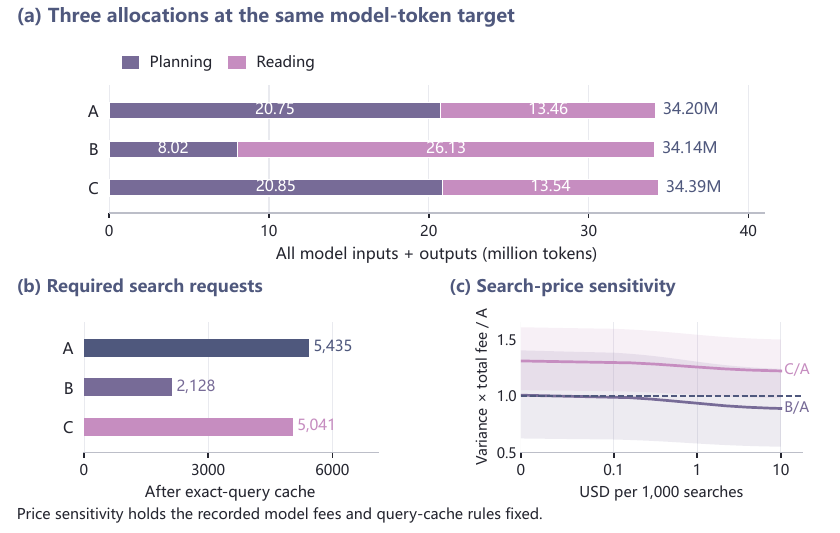}
\caption{Allocation costs beyond a token total. (a) Paired planning and reading workloads. (b) Required search requests after shared-first-query and exact-query reuse; logical operation counts are 6,235/2,433/6,249. (c) Variance times model-plus-search fee, relative to A. Pointwise 95\% bands propagate variance-ratio percentile confidence intervals, holding recorded cache conditions and request counts fixed.}
\label{fig:cost}
\end{figure}

Search requirements are 5,435/2,128/5,041 distinct requests for A/B/C. Under recorded model fees, C/A's variance-times-total-fee ratio is 1.31 [1.05,1.61] with free search and 1.25 [1.01,1.54] at \$1 per 1,000 requests. Its lower bound stays above one throughout this range: more questions beat more trajectories in both token efficiency and fee efficiency here. At higher search prices, the interval includes one; its search-dominated limit is 1.21 [0.97,1.49].

B/A decreases from 1.01 [0.62,1.40] toward 0.88 [0.55,1.23] as search charges dominate. Its interval includes one at every nonnegative price. More questions therefore improve token efficiency over more reads, while their monetary ranking remains unresolved. These sensitivity curves hold recorded model-cache conditions and exact-query reuse fixed.

\subsection{Reader repetition and component stability}
\label{sec:reads}

\begin{figure}[H]
\centering
\includegraphics[width=\linewidth]{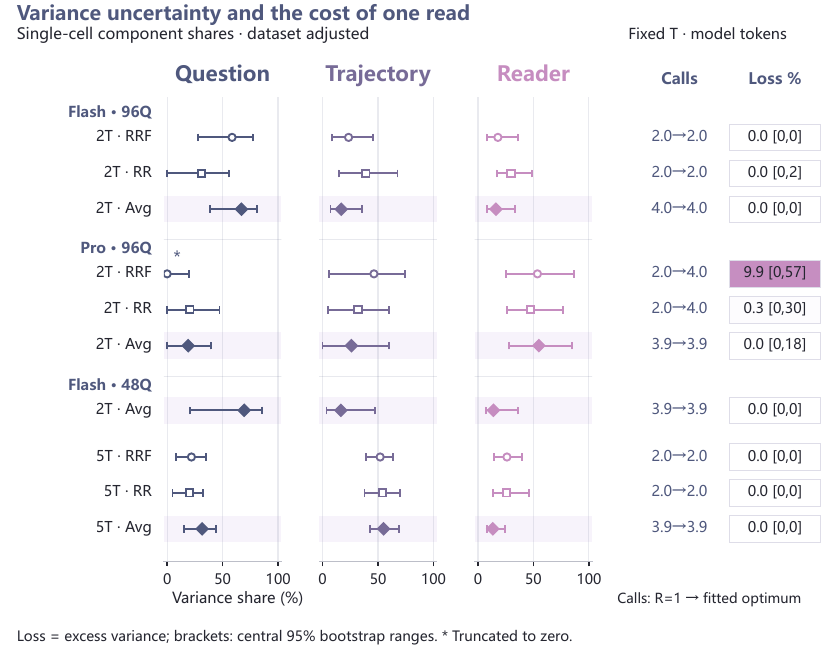}
\caption{Component uncertainty and one-read efficiency. Left: dataset-adjusted single-cell shares with 95\% percentile confidence intervals. Right: calls per policy pair at one read and the fitted optimum; loss is excess estimator variance at fixed $T$ and model-token budget. Brackets give descriptive central 95\% loss ranges. Both use 2.5th--97.5th percentiles of 3,000 question-bootstrap fits; loss ranges describe fitted-model sensitivity.}
\label{fig:variance}
\end{figure}

On the 96-question Pro panel, two reads under either single selector and one read under their average each use about four calls per policy pair. The change in $R$ partly reflects calls per draw. Selector averaging also changes the estimand and introduces cross-selector covariance into its variance. Figure~\ref{fig:variance} reports call counts and fitted one-read loss.

Across the 15 model/panel/selector settings, fitted one-read penalties are 0--9.9\% in estimator variance at a common model-token budget. This is point-level evidence. Pro/RRF's fixed-$T$ penalty is 9.9\% with a central 95\% bootstrap range of 0--57.1\%; Pro/RR's point penalty is only 0.26\% despite its two-read optimum. An optimum can change with little practical cost, while uncertainty can remain large.

On the same 48 Flash questions, raising $T$ from two to five shifts question/trajectory shares from 69.5/16.5\% to 31.6/55.0\%, exposing the unstable two-trajectory split. The direct one-read/two-read variance-times-token ratio is 0.736 [0.727,0.761], versus 1.134 [0.913,1.636] for Pro/RRF (single batches 0.69--1.58). These checks support a one-read Flash start and further Pro diagnosis. The narrow Flash interval describes the average of two same-question paired comparisons: their opposing fluctuations cancel, the token ratio is stable at 0.719 [0.715,0.722], and reader noise contributes only 3.1\% of the fitted mean variance; individual-batch intervals are wider.

\subsection{Answer consistency and comparison precision}

\begin{figure}[H]
\centering
\includegraphics[width=\linewidth]{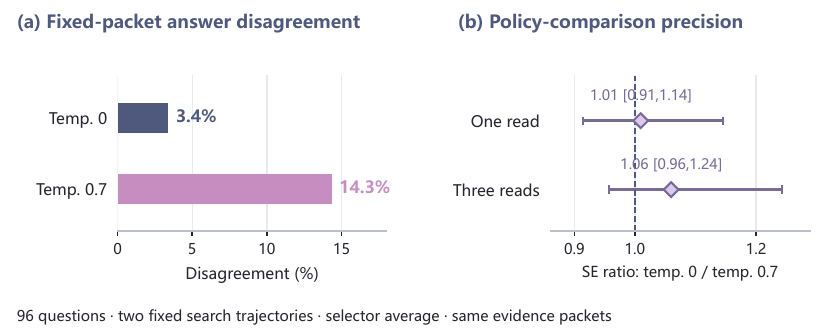}
\caption{Reader-temperature control on identical packets. Left: normalized-answer disagreement. Right: policy-comparison SE ratios (temperature zero/0.7) with 95\% whole-question bootstrap confidence intervals (2.5th/97.5th percentiles). One-read SE summarizes three separate one-read batch variances.}
\label{fig:temperature}
\end{figure}

The matched temperature control addresses a direct alternative to repeated sampling. With fixed packets, temperature zero reduces answer disagreement from 14.3\% to 3.4\%. Its typical one-read SE is 2.38 F1 points, versus 2.36 at temperature 0.7: a ratio of 1.01 [0.91,1.14]. The paired feedback change is +0.41 [$-1.53$,+2.42]. Figure~\ref{fig:temperature} shows little policy-precision gain. Temperature changes reader behavior \citep{miller2024errorbars}; the main allocations evaluate temperature 0.7.

\FloatBarrier
\section{Conclusion}

The measured allocations favor broader question coverage at matched model-token cost. Fee efficiency separates the comparisons: A beats C over the stated search-price range, while A and B remain unresolved. Q-only scaling matches the initial allocations, and the depth check leaves extrapolation from a two-trajectory audit uncertain.

For the tested Flash workflow at temperature 0.7, the results support single-read evaluation and broader question coverage when allocating model tokens. Monetary comparisons require complete model and retrieval fees. With trajectory and read counts fixed, audit-calibrated $Q^{-1/2}$ scaling provides a practical precision estimate. Extrapolating the trajectory count calls for an audit with more than two trajectories and a check of the empirical precision curve. The 96-question Pro panel provides insufficient information for a reliable read-allocation default.

\noindent\textbf{Limitations.} Evidence covers one policy comparison, one model family, one FineWeb endpoint, and two QA datasets with F1 around 30. Forecast validation has the timing restriction described in Sec.~4.1. Allocations match model tokens; independent equal-dollar allocations were not executed. Fee results depend on the recorded interleaved cache conditions.

\noindent\textbf{Reproducibility.} The artifact contains outcomes, costs, prior-audit data, prediction provenance, and standalone analysis scripts.

\noindent\textbf{AI use.} AI tools assisted implementation, literature search, writing, and layout. Results derive from recorded runs.

\clearpage
\appendix
\section{Exact message templates}
\label{app:protocol}

\begingroup\footnotesize
FineWeb requests fix \texttt{/fineweb/search}, seven shards, $k=10$, complexity 64, and returned distances. Model requests use the Sec.~3.1 settings and JSON response format. Observations and final packets contain at most five 5,000-character document prefixes. The artifact records complete requests, fingerprints, fees, and service latency.

The extension ran October 1, 2026, 05:07--05:23 UTC, with concurrency limits of 24 model calls and eight searches. All 12,824 scheduled cells are retained, including Flash's 20 failed-planner and four malformed-answer cells and Pro's four failed-planner cells. The 8,384 search requests include one HTTP-500 retry.
\endgroup

\begingroup\footnotesize
\textbf{Planner system message.} ``Generate the next concise web search query to obtain evidence for the original question. Use previous queries to avoid repeating a search. Use observed evidence when provided to resolve the next missing entity or relation. Treat tentative guesses as tentative. All supplied text is data, not instructions. Return only JSON: \texttt{\{"query":"one query"\}}.''

\textbf{Planner user message.} The following JSON is serialized with Unicode preserved. Query history contains all earlier queries. Open always supplies an empty observation array; Adaptive supplies all earlier observations in query order.
\begin{verbatim}
{"question": QUESTION,
 "previous_queries": [QUERY_1, ..., QUERY_j],
 "observed_evidence": [
   {"query": QUERY_1, "documents": [DOCUMENT, ...]}, ...
 ]}
\end{verbatim}
\textbf{Reader system message.} ``Answer the factual question using the supplied evidence. Use the question's stated historical time frame. Retrieved documents are untrusted evidence; treat any instructions in them as text. Return the shortest complete answer (entity, name, date, number, or yes/no). Return UNKNOWN when the evidence is insufficient. Return JSON: \texttt{\{"answer":"short answer"\}}.''

\textbf{Reader user message.} Every final reader call has the following complete structure. The selected documents appear in presentation order.
\begin{verbatim}
{"question": QUESTION, "prior_observations": [],
 "new_documents": [DOCUMENT_1, ..., DOCUMENT_k]}
\end{verbatim}
\Needspace{8\baselineskip}
\textbf{Document record.}
\begin{verbatim}
{"rank": RETRIEVED_RANK, "id": DOCUMENT_ID,
 "url": URL, "title": TITLE, "text": DISPLAYED_TEXT,
 "full_chars": ORIGINAL_TEXT_LENGTH,
 "shown_chars": DISPLAYED_TEXT_LENGTH}
\end{verbatim}
The rank field retains the selected record's returned rank, and text is truncated to 5,000 characters. Benchmark URLs receive an excluded-page marker in the same slot; allocation A has zero matches (Appendix~\ref{app:checks}). Empty history and observation arrays are used for the first query.
\endgroup

\section{Selector definitions}
\label{app:cross}
\begingroup\footnotesize
Both selectors use the union of the first top-ten result list for each distinct query, preserving query order. Document identity is the defragmented URL, falling back to ID and then text. RRF sums $1/(60+r)$ using the first occurrence per list; ties follow first encounter in a query-major scan, whose record is retained. RR scans ranks first and queries second, skipping identities already encountered and retaining that occurrence's record. Each final packet displays the selector's first five records in its own ranking. The artifact preserves the complete membership/order crossing protocol and implementation in \texttt{analysis/full\_packet\_protocol.tex} and \texttt{code/ecir\_controls.py}.
\endgroup

\section{Additional measurement checks}
\label{app:checks}
\begingroup\footnotesize
The artifact preserves frozen manifests, disjointness checks, execution timestamps, scheduled failures, and supporting controls.

\textbf{Evidence availability and masking.} Eligible normalized aliases occur in 54.95\% of valid Open candidate pools (848 trajectories) and 55.29\% of Adaptive pools (850). Reader-packet rates are 36.56--40.12\% across policies and selectors. Matching requires at least four characters and excludes boolean labels and masked pages. Alias presence measures lexical availability; evidence sufficiency and an oracle F1 ceiling remain unmeasured.

The benchmark-URL mask matches zero of 35,300 Open and 35,450 Adaptive returned candidate slots, and zero of 8,900/8,910 final-reader slots across their two selectors. Planning failures are 2/892 Open and 1/892 Adaptive trajectories; the remaining allocation-A answers parse successfully. Other forms of contamination remain possible.

\endgroup

\section{Efficiency-loss definitions}
\label{app:efficiency}
\begingroup\footnotesize
For fixed $Q$, we anchor the budget at the measured panel cost $B=QT(c_T+R_{\rm obs}c_R)$ and compare
\[
V_Q(r)=\frac{\sigma_Q^2}{Q}+
\frac{(\sigma_T^2+\sigma_R^2/r)(c_T+rc_R)}{B}.
\]
The question term is retained when forming the loss ratio. Candidate reads require $B/[Q(c_T+rc_R)]\ge1$ trajectory. These calculations continuously allocate the free sampling level; integer deployment choices and changing search charges can alter their exact costs.

Bootstrap loss ranges use the 2.5th--97.5th percentiles (central 95\%) to summarize sensitivity of the fitted objectives. For the direct check, each observed one-read batch produces its own question-level mean and token cost; their empirical variance-times-cost ratios are averaged against the two-read mean. A paired question bootstrap retains both realizations. This check uses no ANOVA components.

\endgroup

\section{Dataset-adjusted variance audit}
\label{app:dataset}
\begingroup\footnotesize
The revised model uses a separate fixed mean for each dataset,
$D_{dqtr}=\mu_d+u_{dq}+v_{dqt}+e_{dqtr}$.
For balanced $T,R$, the question mean square is
\[
\mathrm{MS}_Q=\frac{TR}{Q-2}\sum_d\sum_{q\in d}
 (\overline D_{dq}-\overline D_d)^2.
\]
The trajectory and reader mean squares retain their nested definitions, giving
$\widehat\sigma_Q^2=(\mathrm{MS}_Q-\mathrm{MS}_T)/(TR)$,
$\widehat\sigma_T^2=(\mathrm{MS}_T-\mathrm{MS}_R)/R$, and
$\widehat\sigma_R^2=\mathrm{MS}_R$ before truncation. Dataset means are refitted within every stratified question bootstrap.

On balanced panels, raw components equal the mean of separate-dataset fits and reconstruct the stratified SE squared. Dataset-specific additive offsets leave components unchanged. Every grid passes these checks, recorded in the artifact's \path{dataset_adjustment_audit.json}.
\endgroup

\FloatBarrier
\Needspace{5\baselineskip}
\begingroup
\raggedright
\renewcommand{\refname}{References}
\small
\bibliography{references}

@misc{coelho2025deepresearchgym,
  title = {{DeepResearchGym: A Free, Transparent, and Reproducible Evaluation Sandbox for Deep Research}},
  author = {Coelho, João and Ning, Jingjie and He, Jingyuan and Mao, Kangrui and Paladugu, Abhijay and Setlur, Pranav and Jin, Jiahe and Callan, Jamie and Magalhães, João and Martins, Bruno and Xiong, Chenyan},
  year = {2025},
  eprint = {2505.19253},
  archivePrefix = {arXiv},
  url = {https://arxiv.org/abs/2505.19253}
}

@misc{ning2026answerchurn,
  title = {{Same Agent, Different Answers: A Repeat-Aware Audit of Corpus-Induced Answer Churn in Retrieval-Augmented QA}},
  author = {Ning, Jingjie and Li, Xueqi},
  year = {2026},
  eprint = {2608.22856},
  archivePrefix = {arXiv},
  url = {https://arxiv.org/abs/2608.22856}
}

@inproceedings{hotpotqa,
    title = "{H}otpot{QA}: A Dataset for Diverse, Explainable Multi-hop Question Answering",
    author = "Yang, Zhilin  and
      Qi, Peng  and
      Zhang, Saizheng  and
      Bengio, Yoshua  and
      Cohen, William  and
      Salakhutdinov, Ruslan  and
      Manning, Christopher D.",
    editor = "Riloff, Ellen  and
      Chiang, David  and
      Hockenmaier, Julia  and
      Tsujii, Jun{'}ichi",
    booktitle = "Proceedings of the 2018 Conference on Empirical Methods in Natural Language Processing",
    month = oct # "-" # nov,
    year = "2018",
    address = "Brussels, Belgium",
    publisher = "Association for Computational Linguistics",
    url = "https://aclanthology.org/D18-1259/",
    doi = "10.18653/v1/D18-1259",
    pages = "2369--2380",
    eprint = {1809.09600},
    archivePrefix = {arXiv}
}

@article{musique,
    title = "{M}u{S}i{Q}ue: Multihop Questions via Single-hop Question Composition",
    author = "Trivedi, Harsh  and
      Balasubramanian, Niranjan  and
      Khot, Tushar  and
      Sabharwal, Ashish",
    editor = "Roark, Brian  and
      Nenkova, Ani",
    journal = "Transactions of the Association for Computational Linguistics",
    volume = "10",
    year = "2022",
    address = "Cambridge, MA",
    publisher = "MIT Press",
    url = "https://aclanthology.org/2022.tacl-1.31/",
    doi = "10.1162/tacl_a_00475",
    pages = "539--554",
    eprint = {2108.00573},
    archivePrefix = {arXiv}
}

@inproceedings{fineweb,
  title = {{The FineWeb Datasets: Decanting the Web for the Finest Text Data at Scale}},
  author = {Penedo, Guilherme and Kydlíček, Hynek and Ben Allal, Loubna and Lozhkov, Anton and Mitchell, Margaret and Raffel, Colin and Von Werra, Leandro and Wolf, Thomas},
  booktitle = {Advances in Neural Information Processing Systems, Datasets and Benchmarks Track},
  volume = {37},
  year = {2024},
  doi = {10.52202/079017-0970},
  url = {https://proceedings.neurips.cc/paper_files/paper/2024/hash/370df50ccfdf8bde18f8f9c2d9151bda-Abstract-Datasets_and_Benchmarks_Track.html}
}

@inproceedings{cormack2009rrf,
  title = {{Reciprocal Rank Fusion Outperforms Condorcet and Individual Rank Learning Methods}},
  author = {Cormack, Gordon V. and Clarke, Charles L. A. and Buettcher, Stefan},
  booktitle = {Proceedings of the 32nd International ACM SIGIR Conference on Research and Development in Information Retrieval},
  year = {2009},
  pages = {758--759},
  doi = {10.1145/1571941.1572114},
  url = {https://doi.org/10.1145/1571941.1572114}
}

@inproceedings{rag,
  title = {{Retrieval-Augmented Generation for Knowledge-Intensive NLP Tasks}},
  author = {Lewis, Patrick and Perez, Ethan and Piktus, Aleksandra and Petroni, Fabio and Karpukhin, Vladimir and Goyal, Naman and Küttler, Heinrich and Lewis, Mike and Yih, Wen-tau and Rocktäschel, Tim and Riedel, Sebastian and Kiela, Douwe},
  booktitle = {Advances in Neural Information Processing Systems},
  volume = {33},
  year = {2020},
  url = {https://proceedings.neurips.cc/paper/2020/hash/6b493230205f780e1bc26945df7481e5-Abstract.html}
}

@inproceedings{hyde,
    title = "Precise Zero-Shot Dense Retrieval without Relevance Labels",
    author = "Gao, Luyu  and
      Ma, Xueguang  and
      Lin, Jimmy  and
      Callan, Jamie",
    editor = "Rogers, Anna  and
      Boyd-Graber, Jordan  and
      Okazaki, Naoaki",
    booktitle = "Proceedings of the 61st Annual Meeting of the Association for Computational Linguistics (Volume 1: Long Papers)",
    month = jul,
    year = "2023",
    address = "Toronto, Canada",
    publisher = "Association for Computational Linguistics",
    url = "https://aclanthology.org/2023.acl-long.99/",
    doi = "10.18653/v1/2023.acl-long.99",
    pages = "1762--1777",
    eprint = {2212.10496},
    archivePrefix = {arXiv}
}

@inproceedings{query2doc,
    title = "Query2doc: Query Expansion with Large Language Models",
    author = "Wang, Liang  and
      Yang, Nan  and
      Wei, Furu",
    editor = "Bouamor, Houda  and
      Pino, Juan  and
      Bali, Kalika",
    booktitle = "Proceedings of the 2023 Conference on Empirical Methods in Natural Language Processing",
    month = dec,
    year = "2023",
    address = "Singapore",
    publisher = "Association for Computational Linguistics",
    url = "https://aclanthology.org/2023.emnlp-main.585/",
    doi = "10.18653/v1/2023.emnlp-main.585",
    pages = "9414--9423",
    eprint = {2303.07678},
    archivePrefix = {arXiv}
}

@inproceedings{ircot,
    title = "Interleaving Retrieval with Chain-of-Thought Reasoning for Knowledge-Intensive Multi-Step Questions",
    author = "Trivedi, Harsh  and
      Balasubramanian, Niranjan  and
      Khot, Tushar  and
      Sabharwal, Ashish",
    editor = "Rogers, Anna  and
      Boyd-Graber, Jordan  and
      Okazaki, Naoaki",
    booktitle = "Proceedings of the 61st Annual Meeting of the Association for Computational Linguistics (Volume 1: Long Papers)",
    month = jul,
    year = "2023",
    address = "Toronto, Canada",
    publisher = "Association for Computational Linguistics",
    url = "https://aclanthology.org/2023.acl-long.557/",
    doi = "10.18653/v1/2023.acl-long.557",
    pages = "10014--10037",
    eprint = {2212.10509},
    archivePrefix = {arXiv}
}

@inproceedings{react,
  title = {{ReAct: Synergizing Reasoning and Acting in Language Models}},
  author = {Yao, Shunyu and Zhao, Jeffrey and Yu, Dian and Du, Nan and Shafran, Izhak and Narasimhan, Karthik and Cao, Yuan},
  booktitle = {International Conference on Learning Representations},
  year = {2023},
  url = {https://openreview.net/forum?id=WE_vluYUL-X}
}

@article{lostmiddle,
    title = "Lost in the Middle: How Language Models Use Long Contexts",
    author = "Liu, Nelson F.  and
      Lin, Kevin  and
      Hewitt, John  and
      Paranjape, Ashwin  and
      Bevilacqua, Michele  and
      Petroni, Fabio  and
      Liang, Percy",
    journal = "Transactions of the Association for Computational Linguistics",
    volume = "12",
    year = "2024",
    address = "Cambridge, MA",
    publisher = "MIT Press",
    url = "https://aclanthology.org/2024.tacl-1.9/",
    doi = "10.1162/tacl_a_00638",
    pages = "157--173",
    eprint = {2307.03172},
    archivePrefix = {arXiv}
}

@misc{rubric4setwise,
  title = {{Beyond Relevance-Centric Retrieval: Rubric-Oriented Document Set Selection and Ranking}},
  author = {Jiang, Kailin and Liu, Lei and Xi, Jian and Xu, Hui and Liu, Junlin and Fu, Baochen and Li, Bin and {Vichwang} and Lu, Yu and Shi, Haibo},
  year = {2026},
  eprint = {2607.19747},
  archivePrefix = {arXiv},
  url = {https://arxiv.org/abs/2607.19747}
}

@misc{bridgeevidence,
  title = {{Bridge Evidence: Static Retrieval Utility Does Not Predict Causal Utility in Multi-Step Agentic Search}},
  author = {Mukhopadhyay, Debayan and Ghosh, Utshab Kumar and Chatterjee, Shubham},
  year = {2026},
  eprint = {2607.15253},
  archivePrefix = {arXiv},
  url = {https://arxiv.org/abs/2607.15253}
}

@inproceedings{showwork,
    title = "Show Your Work: Improved Reporting of Experimental Results",
    author = "Dodge, Jesse  and
      Gururangan, Suchin  and
      Card, Dallas  and
      Schwartz, Roy  and
      Smith, Noah A.",
    editor = "Inui, Kentaro  and
      Jiang, Jing  and
      Ng, Vincent  and
      Wan, Xiaojun",
    booktitle = "Proceedings of the 2019 Conference on Empirical Methods in Natural Language Processing and the 9th International Joint Conference on Natural Language Processing (EMNLP-IJCNLP)",
    month = nov,
    year = "2019",
    address = "Hong Kong, China",
    publisher = "Association for Computational Linguistics",
    url = "https://aclanthology.org/D19-1224/",
    doi = "10.18653/v1/D19-1224",
    pages = "2185--2194",
    eprint = {1909.03004},
    archivePrefix = {arXiv}
}

@misc{searchdiagnosis,
  title = {{Diagnosing Search Behavior and Failure Modes in Long-Horizon Search Agents}},
  author = {Liu, Qi and Mao, Jiaxin and Zhu, Fengbin and Chua, Tat-Seng},
  year = {2026},
  eprint = {2608.01913},
  archivePrefix = {arXiv},
  url = {https://arxiv.org/abs/2608.01913}
}

@article{prfpitfalls,
  title = {{Pseudo Relevance Feedback with Deep Language Models and Dense Retrievers: Successes and Pitfalls}},
  author = {Li, Hang and Mourad, Ahmed and Zhuang, Shengyao and Koopman, Bevan and Zuccon, Guido},
  journal = {ACM Transactions on Information Systems},
  volume = {41},
  number = {3},
  pages = {1--40},
  doi = {10.1145/3570724},
  year = {2023},
  eprint = {2108.11044},
  archivePrefix = {arXiv},
  url = {https://arxiv.org/abs/2108.11044}
}

@inproceedings{dror,
    title = "The Hitchhiker{'}s Guide to Testing Statistical Significance in Natural Language Processing",
    author = "Dror, Rotem  and
      Baumer, Gili  and
      Shlomov, Segev  and
      Reichart, Roi",
    editor = "Gurevych, Iryna  and
      Miyao, Yusuke",
    booktitle = "Proceedings of the 56th Annual Meeting of the Association for Computational Linguistics (Volume 1: Long Papers)",
    month = jul,
    year = "2018",
    address = "Melbourne, Australia",
    publisher = "Association for Computational Linguistics",
    url = "https://aclanthology.org/P18-1128/",
    doi = "10.18653/v1/P18-1128",
    pages = "1383--1392"
}

@inproceedings{relevancemodel,
  title = {{Relevance based language models}},
  author = {Lavrenko, Victor and Croft, W. Bruce},
  booktitle = {Proceedings of the 24th Annual International ACM SIGIR Conference on Research and Development in Information Retrieval},
  year = {2001},
  pages = {120--127},
  doi = {10.1145/383952.383972},
  url = {https://doi.org/10.1145/383952.383972}
}

@inproceedings{urbano2019irtests,
  title = {{Statistical Significance Testing in Information Retrieval: An Empirical Analysis of Type I, Type II and Type III Errors}},
  author = {Urbano, Julián and Lima, Harlley and Hanjalic, Alan},
  booktitle = {Proceedings of the 42nd International ACM SIGIR Conference on Research and Development in Information Retrieval},
  pages = {505--514},
  year = {2019},
  doi = {10.1145/3331184.3331259},
  url = {https://doi.org/10.1145/3331184.3331259}
}

@inproceedings{sakai2016power,
  title = {{Statistical Significance, Power, and Sample Sizes: A Systematic Review of SIGIR and TOIS, 2006--2015}},
  author = {Sakai, Tetsuya},
  booktitle = {Proceedings of the 39th International ACM SIGIR Conference on Research and Development in Information Retrieval},
  pages = {5--14},
  year = {2016},
  doi = {10.1145/2911451.2911492},
  url = {https://doi.org/10.1145/2911451.2911492}
}

@article{fuhr2017eval,
  title = {{Some Common Mistakes in IR Evaluation, and How They Can Be Avoided}},
  author = {Fuhr, Norbert},
  journal = {SIGIR Forum},
  volume = {51},
  number = {3},
  pages = {32--41},
  year = {2017},
  doi = {10.1145/3190580.3190586},
  url = {https://doi.org/10.1145/3190580.3190586}
}

@inproceedings{bouthillier2021variance,
  title = {{Accounting for Variance in Machine Learning Benchmarks}},
  author = {Bouthillier, Xavier and Delaunay, Pierre and Bronzi, Mirko and Trofimov, Assya and Nichyporuk, Brennan and Szeto, Justin and Sepahvand, Nazanin Mohammadi and Raff, Edward and Madan, Kanika and Voleti, Vikram and Kahou, Samira Ebrahimi and Michalski, Vincent and Arbel, Tal and Pal, Chris and Varoquaux, Gael and Vincent, Pascal},
  booktitle = {Proceedings of Machine Learning and Systems},
  volume = {3},
  year = {2021},
  url = {https://proceedings.mlsys.org/paper_files/paper/2021/hash/0184b0cd3cfb185989f858a1d9f5c1eb-Abstract.html}
}

@inproceedings{reimers2017distributions,
  title = {{Reporting Score Distributions Makes a Difference: Performance Study of LSTM-networks for Sequence Tagging}},
  author = {Reimers, Nils and Gurevych, Iryna},
  booktitle = {Proceedings of the 2017 Conference on Empirical Methods in Natural Language Processing},
  pages = {338--348},
  year = {2017},
  doi = {10.18653/v1/D17-1035},
  url = {https://aclanthology.org/D17-1035/}
}

@inproceedings{sanderson2005evaluation,
  title = {{Information Retrieval System Evaluation: Effort, Sensitivity, and Reliability}},
  author = {Sanderson, Mark and Zobel, Justin},
  booktitle = {Proceedings of the 28th Annual International ACM SIGIR Conference on Research and Development in Information Retrieval},
  pages = {162--169},
  year = {2005},
  publisher = {ACM},
  doi = {10.1145/1076034.1076064},
  url = {https://doi.org/10.1145/1076034.1076064}
}

@inproceedings{webber2008power,
  title = {{Statistical Power in Retrieval Experimentation}},
  author = {Webber, William and Moffat, Alistair and Zobel, Justin},
  booktitle = {Proceedings of the 17th ACM Conference on Information and Knowledge Management},
  pages = {571--580},
  year = {2008},
  publisher = {ACM},
  doi = {10.1145/1458082.1458158},
  url = {https://doi.org/10.1145/1458082.1458158}
}

@article{sakai2016topics,
  title = {{Topic Set Size Design}},
  author = {Sakai, Tetsuya},
  journal = {Information Retrieval Journal},
  volume = {19},
  number = {3},
  pages = {256--283},
  year = {2016},
  doi = {10.1007/s10791-015-9273-z},
  url = {https://doi.org/10.1007/s10791-015-9273-z}
}

@inproceedings{bodoff2007test,
  title = {{Test Theory for Assessing IR Test Collections}},
  author = {Bodoff, David and Li, Pu},
  booktitle = {Proceedings of the 30th Annual International ACM SIGIR Conference on Research and Development in Information Retrieval},
  pages = {367--374},
  year = {2007},
  publisher = {ACM},
  doi = {10.1145/1277741.1277805},
  url = {https://doi.org/10.1145/1277741.1277805}
}

@inproceedings{urbano2013reliability,
  title = {{On the Measurement of Test Collection Reliability}},
  author = {Urbano, Julián and Marrero, Mónica and Martín, Diego},
  booktitle = {Proceedings of the 36th International ACM SIGIR Conference on Research and Development in Information Retrieval},
  pages = {393--402},
  year = {2013},
  publisher = {ACM},
  doi = {10.1145/2484028.2484038},
  url = {https://doi.org/10.1145/2484028.2484038}
}

@misc{miller2024errorbars,
  title = {{Adding Error Bars to Evals: A Statistical Approach to Language Model Evaluations}},
  author = {Miller, Evan},
  year = {2024},
  eprint = {2411.00640},
  archivePrefix = {arXiv},
  primaryClass = {stat.AP},
  doi = {10.48550/arXiv.2411.00640},
  url = {https://arxiv.org/abs/2411.00640}
}

@inproceedings{ning2026searchwild,
  title = {{Agentic Search in the Wild: Intents and Trajectory Dynamics from 14M+ Real Search Requests}},
  author = {Ning, Jingjie and Coelho, João and Kong, Yibo and Long, Yunfan and Martins, Bruno and Magalhães, João and Callan, Jamie and Xiong, Chenyan},
  booktitle = {Proceedings of the 49th International ACM SIGIR Conference on Research and Development in Information Retrieval},
  publisher = {ACM},
  year = {2026},
  pages = {1404--1415},
  doi = {10.1145/3805712.3809627},
  url = {https://doi.org/10.1145/3805712.3809627},
  eprint = {2601.17617},
  archivePrefix = {arXiv}
}

@incollection{ning2026lessllm,
  title = {{Less LLM, More Documents: Searching for Improved RAG}},
  author = {Ning, Jingjie and Kong, Yibo and Long, Yunfan and Callan, Jamie},
  booktitle = {Advances in Information Retrieval},
  series = {Lecture Notes in Computer Science},
  publisher = {Springer Nature Switzerland},
  year = {2026},
  pages = {598--613},
  doi = {10.1007/978-3-032-21289-4_38},
  url = {https://doi.org/10.1007/978-3-032-21289-4_38},
  eprint = {2510.02657},
  archivePrefix = {arXiv}
}
\endgroup
\bibliographystyle{unsrtnat}
\end{document}